# Superconductivity in $SrB_3C_3$ clathrate

Li Zhu[1], Hanyu Liu[1,2], Maddury Somayazulu[3], Yue Meng[3], Piotr A. Guńka[1,4], Thomas B. Shiell[1], Curtis Kenney-Benson[3], Stella Chariton,[5] Vitali B. Prakapenka,[5] Hyeok Yoon,[6] Jarryd A. Horn,[6] Johnpierre Paglione,[6,7] Roald Hoffmann[8], R. E. Cohen[1] and Timothy A. Strobel[1,*]

*[1]Earth and Planets Laboratory, Carnegie Institution for Science, 5251 Broad Branch Road, NW, Washington, DC 20015, USA*

*[2]International Center for Computational Method and Software and State Key Laboratory of Superhard Materials, College of Physics, Jilin University, Changchun 130012, China*

*[3]HPCAT, X-ray Science Division, Argonne National Laboratory, Lemont, Illinois 60439, USA*

*[4]Faculty of Chemistry, Warsaw University of Technology, Noakowskiego 3, 00-664 Warszawa, Poland*

*[5]Center for Advanced Radiation Sources, University of Chicago, Chicago, Illinois 60439, USA*

*[6]Maryland Quantum Materials Center, Department of Physics, University of Maryland, College Park, MD, USA*

*[7]The Canadian Institute for Advanced Research, Toronto, Ontario M5G 1Z8, Canada*

*[8]Department of Chemistry and Chemical Biology, Baker Laboratory, Cornell University, Ithaca, NY 14853-1301, USA*

We predict superconductivity for the carbon–boron clathrate $SrB_3C_3$ at 27–43 K for Coulomb pseudopotential ($\mu^*$) values between 0.17 and 0.10 using first-principles calculations with conventional electron–phonon coupling. Electrical transport measurements, facilitated by a novel *in situ* experimental design compatible with extreme synthesis conditions (>3000 K at 50 GPa), show non-hysteretic resistivity drops that track the calculated magnitude and pressure dependence of superconductivity for $\mu^* \approx 0.15$, and transport measurements collected under applied magnetic fields confirm superconductivity with an onset $T_c$ of approximately 20 K at 40 GPa. Carbon-based clathrates thus represent a new class of superconductors similar to other covalent metals like $MgB_2$ and doped fullerenes. Carbon clathrates share structures similar to superconducting superhydrides, but covalent C–B bonds allow metastable persistence at ambient conditions.

Since superconductivity was first discovered in elemental mercury [1], the search for superconductors with high transition temperatures ($T_c$) has represented an active area of research for more than a century. While much progress has been made in the field of unconventional superconductors (*i.e.,* cuprates [2,3] and iron-pnictides [4–6]), the underlying mechanisms for these materials remain controversial, which creates challenges for the design of new higher-$T_c$ materials. On the other hand, recent high-pressure studies on hydride materials [7–20] have pushed superconductivity closer to room temperature. The superconducting mechanisms of these hydrides appear to be well-described by the Eliashberg theory of phonon-mediated superconductivity, and density functional theory (DFT) predicts their electronic structures, phonons, and electron–phonon coupling quite well [21,22]. In this case, high-frequency phonons account for a large portion of the high superconducting transition temperature, thus, materials containing light elements have been considered as optimal candidates for high-$T_c$ superconductivity.

As the lightest element, metallic hydrogen was first considered to be a high-temperature superconductor [23], followed by several high-$T_c$ (super)hydrides based on the concept of chemical pre-compression [7–20,24]. Current evidence suggests that these materials [7–20,24] are electron–phonon superconductors, and that DFT computations may be used to guide experiments. Indeed, $H_3S$ [8,25–27], $LaH_{10}$ [11,13,16,28,29] and others [30–35] were predicted to be high-$T_c$ superconductors before experiments were performed. Recent predictions even suggest the possibility for above-room-temperature superconductivity in $Li_2MgH_{16}$ [12]. Most of the high-$T_c$ superhydrides adopt clathrate-like structures in which H atomic positions represent the vertices of polyhedral cages and weak bonds between H atoms represent cage edges, similar to tetrahedral clathrate host lattices [36–41]. To date, all known superhydride materials decompose at low pressures (most superhydrides are formed above 150 GPa). Stabilization of superhydride superconductors at ambient conditions presents a major challenge, and an open fundamental question is whether other light-element compounds can approach the high-$T_c$ values observed for superhydrides, but at reduced pressures.

Other light-element compounds aside from hydrides also exhibit superconductivity and possess strong covalent bonds that enable their persistence at ambient pressure. For example, $MgB_2$ is the highest-temperature conventional superconductor known at atmospheric pressure with $T_c$ = 39 K [42]. Structurally similar to $MgB_2$, superconductivity ($T_c$ < 5 K) was demonstrated for $MB_2C_2$ (M = rare-earth metal) compounds, and higher $T_c$'s have been predicted for $MB_2C_2$ compounds containing alkali and alkaline-earth metals [43–45]. Doped diamond [46–48] and fullerenes [49] also exhibit moderately high $T_c$. Recently, superconductivity was predicted in undoped diamond via compression-shear deformation [50] and boron-carbon compounds [51]. Carbon-based clathrates, sharing similar structures to the high-pressure superhydrides, were predicted to display high-$T_c$ superconductivity at atmospheric pressure. For example, $T_c \approx 119$ K was predicted for Na-filled carbon clathrate in the sodalite structure [52], and $T_c \approx 77$ K was predicted for Type-II carbon clathrate doped with fluorine guest atoms [53]. In these cases, $sp^3$ bonding may facilitate large electron–phonon coupling [54]. Nevertheless, clathrate structures represent challenging synthetic targets due to large formation energies.

Using a combination of swarm-intelligence-based structure prediction methods [55–57] and high-pressure experiments, we recently predicted and synthesized the first carbon-based clathrate, $SrB_3C_3$ [Fig. 1(a)], which is stabilized by boron substitution into the host framework [58]. $SrB_3C_3$ is metallic with an appreciable density of states at the Fermi energy, suggesting the potential for superconductivity at moderately high temperature. In this Letter, we predict superconductivity for $SrB_3C_3$ using first-principles electron–phonon calculations and report experimental electrical transport measurements that are consistent with the calculations. Our results suggest that the $SrB_3C_3$ clathrate exhibits a moderately high $T_c$ of ~22 K at 23 GPa, which is estimated to increase to ~31 K at ambient pressure.

We computed the electronic band structure $SrB_3C_3$ clathrate using the SCAN meta generalized gradient approximation functional [59], which was demonstrated to produce accurate electronic structures [60]. The electron deficiency of the B–C framework creates empty bands above the Fermi level. Figure 1(b) also shows the band projections onto the B 2*p* and C 2*p* orbitals. Compared to other orbitals, the 2*p* orbitals are the most dominant components of the electronic states near the Fermi level [Fig. 1(c)], indicating that the electrical transport properties are primarily controlled by the B–C framework. The band structure of $SrB_3C_3$ shows that the conduction behavior crosses from hole-like to electron-like near the M point close to the Fermi energy, and several dispersive (steep) bands that cross the Fermi surface along R–M–Γ. These dispersive bands are responsible for electron conduction through the B–C framework. We note also that, as the band structure and DOS show, one could move the Fermi level into regions of a higher DOS by electron doping from the $SrB_3C_3$ stoichiometry.

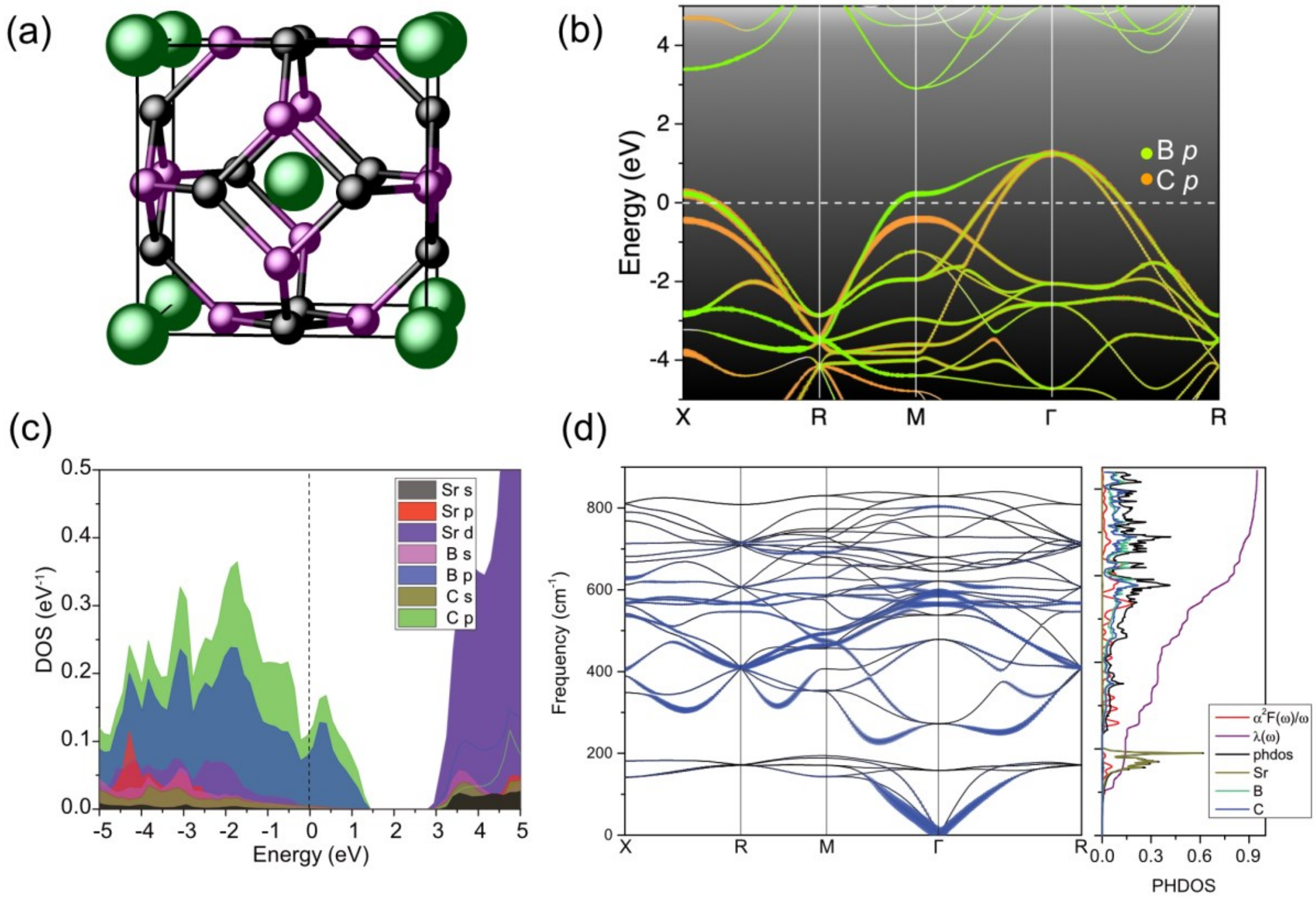

**FIG. 1. (a)** The crystal structure of the $SrB_3C_3$ clathrate. The green, purple and black spheres represent Sr, B, and C atoms, respectively. **(b)** Electronic band structure for $SrB_3C_3$ at 0 GPa. The bands projected onto the B *p* and C *p* orbitals are displayed with green and orange points, respectively. The size of each point is proportional to the weight of the orbital character. **(c)** Projected density of states (DOS) in $SrB_3C_3$. The dashed line indicates the Fermi energy. **(d)** Phonon dispersion, phonon density of states (PHDOS), phonon spectroscopic function $\alpha^2F(\omega)/\omega$ and electron–phonon integral $\lambda(\omega)$ of $SrB_3C_3$ at 0 GPa. The size of blue circles in the phonon dispersion curves is proportional to the electron–phonon coupling strength.

We next performed electron–phonon coupling calculations to investigate potential superconductivity in $SrB_3C_3$ [61]. Phonon dispersion relations and the phonon density of states (PHDOS) for $SrB_3C_3$ clathrate at 0 GPa are shown in Fig 1(d). To illustrate the contributions associated with different phonon modes, we plot blue circles with radii proportional to the corresponding electron–phonon coupling strength, λ. The phonon spectrum is divided into two regions by a gap at 200 cm$^{-1}$. The lower-frequency branch is mainly associated with phonons involving Sr atoms, while the higher frequency branch is mainly associated with phonons involving C and B atoms. The superconducting transition temperature, $T_c$, was estimated from the Allen-Dynes modified McMillan equation [62]. At ambient pressure, the calculated $T_c$ range is 27–43 K for Coulomb pseudopotential values of $\mu^*$ = 0.17–0.10, respectively.

According to the McMillan equation, $T_c$ is dominated by λ, the logarithmic average of the phonon spectrum ($\omega_{log}$), and the DOS at the Fermi level ($N_F$). The electron–phonon coupling parameter λ was evaluated with the McMillan-Hopfield expression [63], from which it is clear that a large $N_F$ has a positive contribution to λ. In the present study, the calculated $N_F$ of $SrB_3C_3$ is 0.13 eV$^{-1}$ per atom at 0 GPa, which is almost completely derived from B–C bonds [Fig. 1(b, c)]. The $N_F$ of $SrB_3C_3$ leads to a large λ of 0.95 compared to the value of 0.71 for $MgB_2$ at ambient pressure [64]. The complete mixing of B–C vibrations throughout their frequency range, except at the highest frequencies (above ~700 cm$^{-1}$), contribute 86% of the total magnitude of λ [Fig. 1(d)]. In contrast, the low-frequency Sr translational vibrations (below 200 cm$^{-1}$) were found to only contribute ~14% in total to λ.

In order to test predictions of superconductivity, we developed a new experimental diamond anvil cell (DAC) technique for *in situ* electrical transport measurements on samples that require extreme synthesis conditions, *e.g.*, 50 GPa and >3000 K. The refractory nature of B and C necessitates prolonged laser heating at very high temperatures, which is not needed in high-pressure hydride experiments that only require gentle pulsed heating at moderate temperatures <1000 K (*e.g.* $LaH_{10}$ [13,29]), or even simple exposure to visible laser irradiation [18]. Establishing a method to produce enough sample to satisfy the electrical transport percolation threshold required to observe superconductivity (~16 vol%) [65] without melting the electrical contacts proved especially challenging. One-sided laser heating with thermal insulation was ineffective to produce enough $SrB_3C_3$ between electrical contacts on the non-heated anvil. Our solution was to use two single crystals of sapphire (~5–10 μm thick) that rest on top of both diamond anvils and provide thermal insulation during heating from a tightly focused infrared laser. In this case, electrical leads are placed on top of one of the sapphire crystals, as shown in Fig. 2. By heating the synthesis precursor in a ~10 μm diameter spot near ~3000 K, we were able to produce samples containing ~60 wt% $SrB_3C_3$ clathrate located between the Pt electrical contacts.

Given the nature of the experimental setup, it is not possible for the electrical probes to directly contact the clathrate phase (which would destroy the probes during heating), and thus additional contact resistance from the precursor phase is always present.

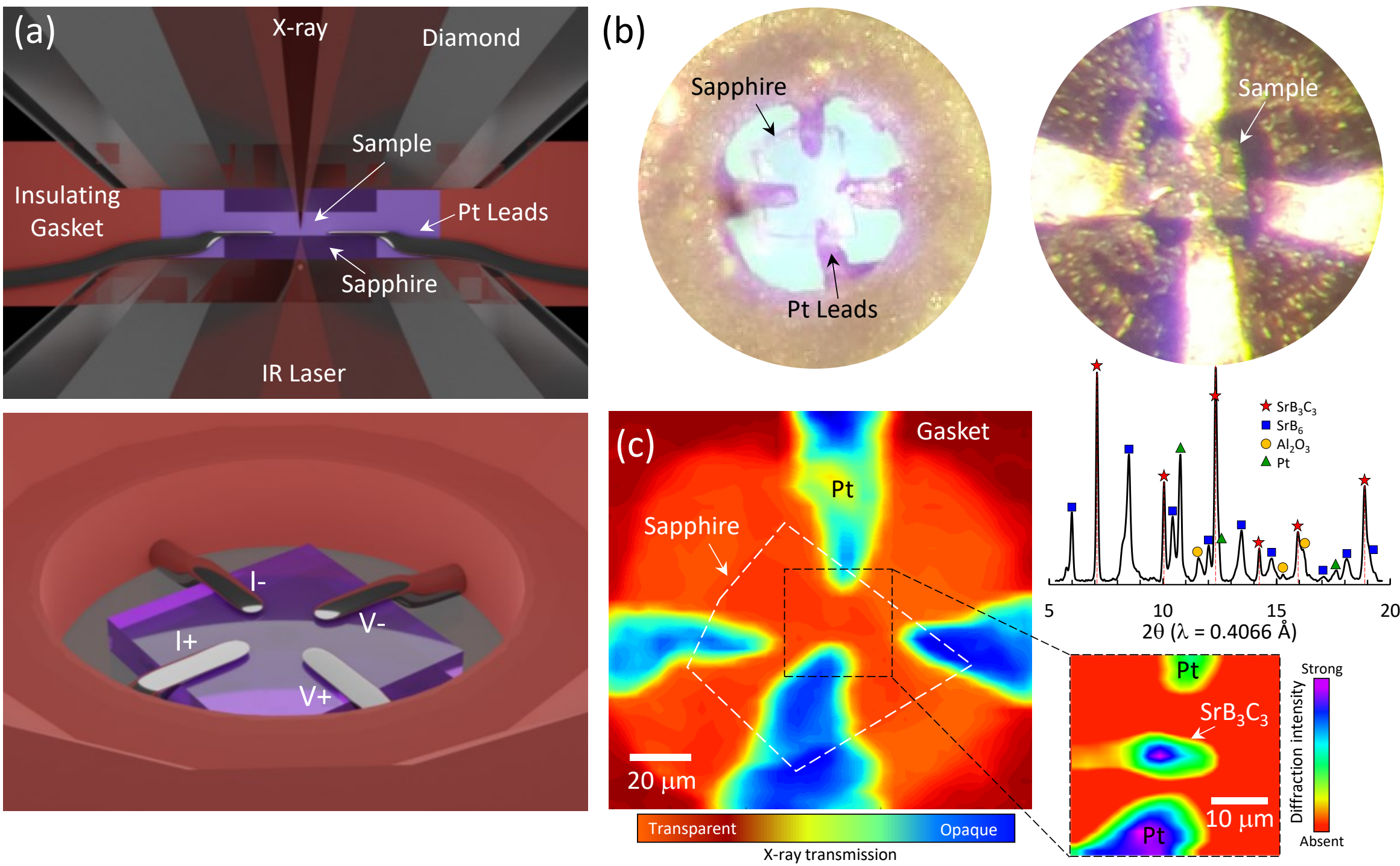

FIG. 2. (a) Sample geometry for *in situ* electrical transport with HPHT synthesis. The 1Sr:3B:3C precursor is pressed between the diamond anvils, separated by sapphire single crystals that serve as thermal insulation. Four Pt electrical probes sit on top of the bottom sapphire crystal and are secured within the insulating gasket. (b) Photomicrographs of the sample chamber before loading (view through top diamond), and after loading sample (view through bottom diamond). (c) X-ray radiograph of sample chamber with four Pt wires. Outline of the sapphire is indicated by the white dashed line. The dashed black square contains the laser heated region between the Pt contacts. The zoom window shows an X-ray diffraction intensity map for $SrB_3C_3$ and Pt showing the synthesized sample between the contacts. A representative XRD pattern from the heated region shows ~60 wt% $SrB_3C_3$ with unconverted cubic $SrB_6$.

After verifying HPHT synthesis of the $SrB_3C_3$ clathrate phase between the electrical leads using synchrotron XRD, the DAC was placed into an open-flow He cryostat equipped for AC electrical transport measurements (see Supplementary Information). The electrical resistivity of the samples initially increases with decreasing temperature (Fig. 3), indicating semiconducting behavior, rather than the metallic behavior expected for the clathrate. This is attributed to incomplete precursor conversion with unreacted glassy carbon, cubic $SrB_6$ and amorphous $SrC_2$, which are all resistive phases and dominate transport at the electrical contacts. Though conducting at ambient pressure, compressed glassy carbon is insulating at 50 GPa [66,67]. Upon cooling $SrB_3C_3$ near ~20 K, the resistance drops sharply by ~30–50%, and then either increases again with the pre-transition slope or continues to fall with additional cooling to the base temperature of ~5 K (Fig. 3). This behavior is consistent with a mixture of superconducting and resistive phases, the precise proportion and position of which determines the total resistance at a given temperature (see Supplementary Information) [68–72]. The change in resistivity is reversible with heating/cooling cycles, showing no detectable hysteresis, consistent with superconductivity, but not with a structural phase transition.

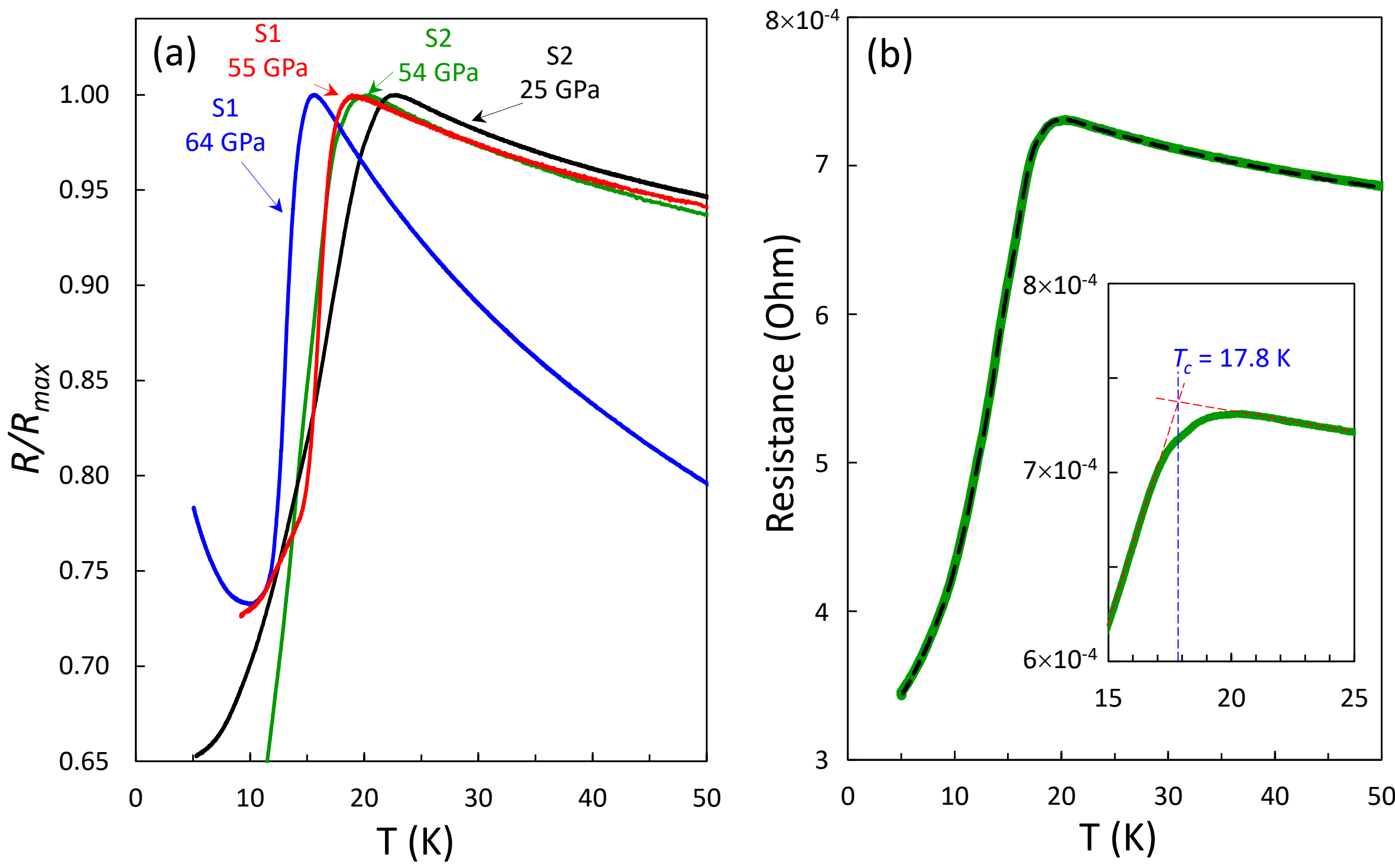

FIG. 3. (a) Normalized resistance for several cooling runs at different pressures for different samples (S1 & S2). (b) Resistance drop during cooling of S2 to a base temperature of 5 K at 54 GPa. The heating trace (dashed line) follows the same path with no detectable hysteresis, as expected for superconductivity. Inset shows the determination of the $T_c$ onset by extrapolation of linear regions.

The $T_c$ onset for different runs was estimated by extrapolating the linear regions before and after the resistance drop. At the highest synthesis pressure of 64 GPa, the experimental $T_c$ = 14.4 K is in good agreement with calculations when $\mu^* \approx 0.15$, which is in the normal range [63]. Previous McMillan fits for silicon clathrate $Ba_8Si_{46}$ [73] and germanium clathrate $Ba_{24}Ge_{100}$ [74,75] gave a range of 0.10–0.31 for $\mu^*$. Our computational predictions indicate that $T_c$ of $SrB_3C_3$ increases with decreasing pressure due to an increase in $N_F$, $\omega_{log}$ and $\lambda$ (Supplementary Information). The experimental resistance drops measured during sample unloading follow the calculated pressure dependence with $\mu^*$ bounded between 0.13–0.17 (Fig. 4).We note that during the preparation of this manuscript, a report of superconductivity in $SrB_3C_3$ using anisotropic Eliashberg equations [76] was published [77], which confirms our calculations with nearly identical values for $T_c$.

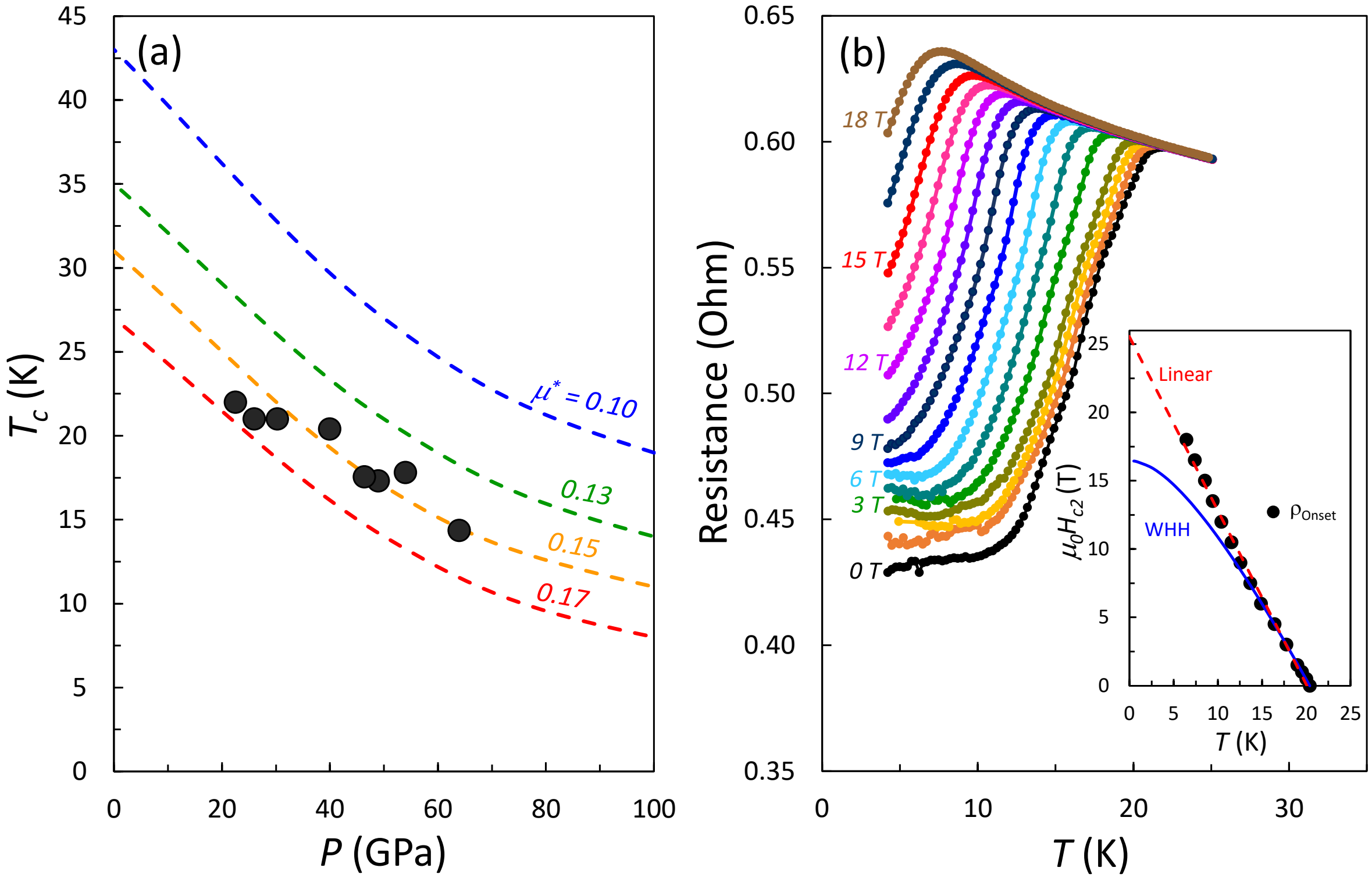

FIG. 4. (a) Experimental $T_c$ onset values for $SrB_3C_3$ with pressure (black points) compared with predictions for different values of $\mu^*$ (dashed lines). (b) Resistance of sample S3 at 40 GPa as a function of temperature and magnetic field. The inset shows $\mu_0 H_{c2}(T)$ for the transition onset (black points) compared with the WHH model ($\alpha$ =0) and linear fit to guide to the eye.

The superconducting nature of the transition was verified by repeating the electrical transport measurements under applied magnetic fields. If the transition is indeed related to superconductivity, the transition temperature will be suppressed up to a critical field, $H_{c2}$, at which point superconductivity will be destroyed and the material will return to its normal conducting state. With increasing field up to 18 T, the transition temperature is indeed suppressed by approximately –0.8 K/T, which confirms the superconducting nature of the sample (Fig. 4). $H_{c2}(T)$ data determined using transition onset criteria at ~40 GPa, are not described well by the Ginzberg–Landau $(1 - t^2)$ model, and deviate significantly from the Werthamer–Helfand–Hohenberg (WHH) equation at fields above ~7 T [78,79]. This deviation, which approximates a linear trend up to 18 T, may have contributions from enhanced electron-phonon coupling ($\lambda_{calc}$ = 0.95) [80], Fermi surface anisotropy, grain effects in the powder sample, or multi-band effects [81]. Similar deviations were observed previously in carbon and boron-bearing superconductors [82–84] and require future investigations.

The agreement with calculations in both the magnitude and pressure dependence of $T_c$, as well suppression of the experimental transition with magnetic field provide compelling evidence that carbon-based clathrates are a promising new class of superconductors. The highest $T_c$ recorded thus far is ~22 K at 23 GPa. Attempts to measure the transition below ~20 GPa were unsuccessful due to experimental complications on decompression such as broken diamonds or loss of electrical contact. Since XRD measurements confirm that $SrB_3C_3$ is recoverable to ambient pressure when held under inert conditions [58], additional future measurements are needed to confirm the presence of superconductivity at 1 atm. Following the theoretical trend for $\mu^*$ = 0.15, $T_c$ near 31 K is anticipated at ambient pressure, which is comparable to the current conventional record.

In summary, we report that carbon clathrates can exhibit high superconducting transition temperatures through a joint theoretical and experimental study. To verify our predictions, we developed a novel experimental design that allows *in situ* four-probe measurements after synthesis at extreme conditions. Calculations indicate that the $T_c$ increases with the decreasing pressure and the experimental results show the same trend. The consistency of the computations and experiments suggests that $SrB_3C_3$ is a conventional superconductor, opening the door to computationally guided optimization of superconducting properties, and paving a novel path for covalent cage materials with high-$T_c$ superconductivity.

**Acknowledgements** We thank R. Hrubiak, M. Guerette, G. M. Borstad, R. Ferry, Z. Geballe, J. Lai and A. Karandikar for assistance with experiments; J. Rojas for assistance with graphics; L. Dubrovinski for providing a non-magnetic cell. This work was supported by the U.S. Department of Energy, Office of Science, Basic Energy Sciences, under Award Number DE-SC0020683. Experiments at the University of Maryland were supported by the Gordon and Betty Moore Foundation's EPiQS Initiative through Grant No. GBMF9071. Computations were carried out at the Memex cluster of Carnegie Institution for Science. Portions of this work were performed at HPCAT (Sector 16), and GSECARS (Sector 13), Advanced Photon Source (APS), Argonne National Laboratory. HPCAT operations are supported by DOE-NNSA's Office of Experimental Sciences. GeoSoilEnviroCARS is supported by the National Science Foundation – Earth Sciences (EAR – 1634415) and Department of Energy- GeoSciences (DE-FG02-94ER14466). The Advanced Photon Source is a U.S. Department of Energy (DOE) Office of Science User Facility operated for the DOE Office of Science by Argonne National Laboratory under Contract No. DE-AC02-06CH11357. P.A.G. thanks Polish National Agency for Academic Exchange for financial support.